\documentclass[fleqn,usenatbib]{mnras}

\usepackage{newtxtext,newtxmath}

\usepackage[T1]{fontenc}

\DeclareRobustCommand{\VAN}[3]{#2}
\let\VANthebibliography\thebibliography
\def\thebibliography{\DeclareRobustCommand{\VAN}[3]{##3}\VANthebibliography}

\usepackage{graphicx}	
\usepackage{amsmath}	
\usepackage{siunitx}

\DeclareSIUnit\parsec{pc}
\DeclareSIUnit\ph{ph}
\DeclareSIUnit\year{yr}
\DeclareSIUnit\simsun{M_\odot}
\DeclareSIUnit\ergs{ergs}
\DeclareSIUnit\ev{eV}
\DeclareSIUnit\byte{B}
\DeclareSIPrefix\comovmega{cM}{6}
\DeclareSIPrefix\propmega{pM}{6}
\DeclareSIPrefix\comovkilo{ck}{3}
\DeclareSIUnit\h{h}
\DeclareSIUnit\angs{\textup{\AA}}
\DeclareSIUnit\atom{H}
\DeclareSIUnit\zsun{Z_\odot}

\newcommand{\codai}{{\fontfamily{pnc}\selectfont {CoDa I}}}
\newcommand{\codaii}{{\fontfamily{pnc}\selectfont {CoDa II}}}
\newcommand{\codaiii}{{\fontfamily{pnc}\selectfont {CoDa III}}}

\newcommand{\mfpacro}{mfp}

\newcommand{\mfp}{$\rm \lambda_{\rm AT}$}
\newcommand{\mfplos}{$\rm \lambda_{\rm los}$}
\newcommand{\mfpion}{$\rm \lambda_{\rm i}$}

\newcommand{\xhi}{$\rm \langle x_{HI} \rangle$}

\newcommand{\gammahi}{$\rm \langle \Gamma_{HII} \rangle$}

\newcommand{\lya}{$\rm Lyman \mhyphen \alpha$}

\mathchardef\mhyphen="2D

\defcitealias{ocvirk_cosmic_2020}{Oc20}
\defcitealias{ocvirk_cosmic_2016}{Oc16}
\defcitealias{ocvirk_lyman-alpha_2021}{Oc21}
\defcitealias{bosman_hydrogen_2022}{Bo22}
\defcitealias{becker_mean_2021}{Be21}
\defcitealias{davies_predicament_2021}{Da21}
\defcitealias{cain_short_2021}{Ca21}
\defcitealias{planck_collaboration_2020}{Planck18}
\defcitealias{collaboration_planck_2014}{Planck14}

\title[Ionising photon mean free path in Cosmic Dawn III]{The short ionizing photon mean free path at z=6 in Cosmic Dawn III, a new fully-coupled radiation-hydrodynamical simulation of the Epoch of Reionization}

\author[Joseph S. W. Lewis]{
Joseph S. W. Lewis$^{1,2}$,
Pierre Ocvirk$^{3}$,
Jenny G. Sorce$^{4,5}$,
Yohan Dubois$^{6}$,
Dominique Aubert$^{3}$,
\newauthor
Luke Conaboy$^{7}$,
Paul R. Shapiro$^{8}$,
Taha Dawoodbhoy$^{8}$,
Romain Teyssier$^{9}$,
Gustavo Yepes$^{10,11}$,
\newauthor
Stefan Gottlöber$^{4}$,
Yann Rasera$^{2}$,
Kyungjin Ahn$^{13}$,
Ilian T. Iliev$^{7}$,
Hyunbae Park$^{14,15,16}$,
\'{E}milie Th\'{e}lie$^{3}$
\\
$^{1}$Zentrum f\"ur Astronomie der Universit\"at Heidelberg, Institut f\"ur Theoretische Astrophysik, Albert-Ueberle-Stra\ss e 2, 69120 Heidelberg, Germany\\
$^{2}$Max-Planck-Institut f\"ur Astronomie, K\"onigstuhl 17, D-69117 Heidelberg, Germany\\
$^{3}$Observatoire Astronomique de Strasbourg, Université de Strasbourg, CNRS UMR 7550, 11 rue de l’Université, 67000 Strasbourg, France\\
$^{4}$Leibniz-Institut für Astrophysik Potsdam (AIP), An der Sternwarte 16, D-14482 Potsdam, Germany\\
$^{5}$Universit\'e Paris-Saclay, CNRS, Institut d'Astrophysique Spatiale, 91405, Orsay, France\\
$^{6}$Institut d’Astrophysique de Paris, UMR 7095, CNRS, UPMC Univ. Paris VI, 98 bis boulevard Arago, 75014 Paris, France\\
$^{7}$Astronomy Center, Department of Physics \& Astronomy, Pevensey II Building, University of Sussex, Falmer, Brighton BN1 9QH, United Kingdom\\
$^{8}$Department of Astronomy, University Texas, Austin, TX 78712-1083, USA\\
$^{9}$Institute for Theoretical Physics, University of Zurich, Winterthurerstrasse 190, CH-8057 Zürich, Switzerland\\
$^{10}$ Departamento de F\'isica Te\'orica M-8, Universidad Aut\'onoma de Madrid, Cantoblanco, 28049, Madrid, Spain\\
$^{11}$ Centro de Investigaci\'on Avanzada en F\'isica  Fundamental (CIAFF), Universidad Aut\'onoma de Madrid, 28049 Madrid, Spain\\
$^{12}$Laboratoire Univers et Th\'eories, Universit\'e de Paris,Observatoire de Paris, Universit\'e PSL, CNRS, F-92190 Meudon, France\\
$^{13}$Chosun University, 375 Seosuk-dong, Dong-gu, Gwangjiu 501-759, Korea\\
$^{14}$Lawrence Berkeley National Laboratory, CA 94720-8139, USA \\
$^{15}$Berkeley Center for Cosmological Physics, UC Berkeley, CA 94720, USA \\
$^{16}$Kavli IPMU (WPI), UTIAS, The University of Tokyo, Kashiwa, Chiba 277-8583, Japan
}
\date{Accepted 2022 August 08.  Received 2022 August 05; in original form 2022 February 11}

\pubyear{2022}

\begin{document}
\label{firstpage}
\pagerange{\pageref{firstpage}--\pageref{lastpage}}
\maketitle

\begin{abstract}
Recent determinations of the mean free path of ionising photons (\mfpacro) in the intergalactic medium (IGM) at $\rm z=6$ are lower than many theoretical predictions. 
In order to gain insight, we investigate the evolution of the \mfpacro \, in our new massive fully coupled radiation hydrodynamics cosmological simulation of reionization: Cosmic Dawn III (\codaiii). 
\codaiii's scale ($\rm 94^3 \, cMpc^3$) and resolution ($\rm 8192^3$ grid) make it particularly suitable to study the IGM during reionization. The simulation was performed with RAMSES-CUDATON on Summit, and used 131072 processors coupled to 24576 GPUs, making it the largest reionization simulation, and largest ever RAMSES simulation. 
A superior agreement with global constraints on reionization is obtained in \codaiii \, over \codaii, especially for the evolution of the neutral hydrogen fraction and the cosmic photo-ionization rate, thanks to an improved calibration, later end of reionization ($\rm z=5.6$), and higher spatial resolution.
Analyzing the \mfpacro, we find that \codaiii \, reproduces the most recent observations very well, from $\rm z=6$ to $\rm z=4.6$. We show that the distribution of the \mfpacro \, in \codaiii \, is bimodal, with short (neutral) and long (ionized) \mfpacro \, modes, due to the patchiness of reionization and the co-existence of neutral versus ionized regions during reionization.
The neutral mode peaks at sub-kpc to kpc scales of \mfpacro, while the ionized mode peak evolves from $\rm 0.1 Mpc/h$ at $\rm z=7$ to $\sim 10$ Mpc/h at $\rm z=5.2$.
Computing the \mfpacro \, as the average of the ionized mode provides the best match to the recent observational determinations. The distribution reduces to a single neutral (ionized) mode at $\rm z>13$ ($\rm z<5$).

\end{abstract}




\section{Introduction}

The Epoch of Reionization (EoR hereafter) began as the first ionising galactic sources started to ionize their immediate surroundings in the inter-galactic medium (IGM) some few hundred million years after the Big Bang. This epoch lasted approximately \SI{600}{\mega\year}, until the reionization of the IGM was complete at some point between $\rm z\approx 6$ and $\rm z \approx 5.3$ \citep[][Bo22 hereafter]{bosman_hydrogen_2022}. Constraining this distant epoch offers unique insight into the formation of the first galaxies and stars, and is an important step in bridging the gap between our knowledge of 'modern' galaxies, and the first moments after the Big Bang. Indeed, the rising and expanding UV background during the EoR may suppress star formation in low mass galaxies  \citep[][the former two will be called Oc16 and Oc20 from hereon]{ocvirk_cosmic_2016,ocvirk_cosmic_2020,dawoodbhoy2018}. This in turn motivated efforts to investigate and predict possible signatures of the EoR in populations of galaxies of the lowest mass, which are best observed in the Local Group \citep[][]{ocvirk2011,iliev_reionization_2011,ocvirk2013,ocvirk2014,dixon_reionization_2018}. 

Due to the very nature of the EoR and the extreme distances involved, direct observations of the reionizing sources as they drive reionization are extremely challenging. Despite this, over the past decades innovative indirect constraints on both the sources of reionization and on the IGM have emerged. Further, the future is bright as upcoming space and ground based observatories (e.g. the James Webb Space Telescope, the Nancy Grace Roman Space Telescope, or the Extremely Large Telescope) will improve our capability to target the drivers of reionization. At the same time, current and planned
21cm radio observations (e.g. EDGES , HERA, LOFAR, NenuFAR, SKA) may bring new constraints on the ionisation of the IGM during the EoR and at cosmic dawn. Thus, the study of the EoR appears to be set for rapid progress over the coming years. In preparation of this new era of EoR observations, the simulation community is hard at work making predictions for these new observational campaigns, and investigating the physical processes that drive the sources of reionization. However, the predictive power of these simulations relies on the quality of their agreement with existing observational constraints, some of which have proved difficult to match.

For instance, recent determinations of the mean free path of ionising photons (\mfpacro \, hereinafter) , at a higher redshift than ever before, have shown that models and simulations of reionization overestimate the \mfpacro \, when $\rm z\approx 6$ \citep[and possibly when $\rm 5.24 \lesssim z \lesssim  6 $, see][]{rahmati_mean_2018, daloisio_hydrodynamic_2020, keating_long_2020}, casting doubt on the realism of their portrayal of the ionisation state of the IGM and thus cosmic reionization as a whole \citep[See][ Be21 from hereon]{becker_mean_2021}. Already, new work has investigated the implications of these new data, finding that they require late and rapid reionization scenarios \citep[][Da21 and Ca21 from hereon]{davies_predicament_2021, cain_short_2021}{, which are also favoured by constraints on the \lya{} forest \citep[][]{kulkarni_large_2019, bosman_hydrogen_2022}}. In fact, \citetalias{cain_short_2021} have been able to obtain a reasonable agreement with both the new and old \mfpacro \, constraints.
We consider below these two recent works attempting to match the new very low \mfpacro \, measurement of \citetalias{becker_mean_2021} and show how using {\fontfamily{pnc}\selectfont {Cosmic Dawn III}} (\codaiii \,)\footnote{More information can be found about the whole simulation suite and the CoDa project  here: \url{https://coda-simulation.github.io/}.} may improve upon them. Both studies above use a fairly low resolution simulation of the IGM ($\rm \sim 1 cMpc$ cell size), and must recourse to complex sub-grid models, to:
\begin{itemize}
    \item model the source population: \citetalias{davies_predicament_2021}'s description of sources involves a range of parameters which fold in a number of complex and coupled aspects, such as the sensitivity of sources to SN and AGN feedback, their star formation efficiency, their escape fractions and the global dependence of these processes with respect to mass.  However, as we show in \cite{ocvirk_lyman-alpha_2021}(Oc21 hereafter), radiative suppression of star formation in low mass haloes is crucial to obtaining a realistic Lyman-alpha forest after overlap. This regulation, for instance, is not included in \citetalias{davies_predicament_2021}, and \codaiii \, improves on their approach by allowing this process to occur self-consistently thanks to the fully coupled radiation hydrodynamics (RHD) framework employed.
    \item set the transmission properties of the IGM sub-grid: \citetalias{cain_short_2021} uses a sub-grid parameterization of the mean free path, which relies on simulations of irradiated patches \citep{daloisio_hydrodynamic_2020}. Such simulations do not include star formation, and one can therefore estimate that their densest patch structures would at some point form stars and therefore, both decrease their own mean free path and the mean free path around them. 

    Our approach circumvents the need for using such a sub-grid model of the mean free path, as the whole volume of \codaiii \, is eligible to star formation provided the relevant criteria in density and temperature are met. Moreover, the \mfpacro \, is self-consistently set in \codaiii \,  by the physical state of the IGM, and is self-consistently affected by radiation received by sources near and far and the high-resolution hydrodynamical description of the IGM.
\end{itemize}
Beyond these two studies, it is desirable, in all cases, to deploy a fully coupled RHD formalism as we do here, in order to break free from the limitations imposed by the parameterization of the physical processes at hand as in \citetalias{davies_predicament_2021,cain_short_2021}, and check that the results they claim can be confirmed in more self-consistent simulations where more physics are resolved and therefore do not require such a level of sub-grid modelling.
In this direction, the THESAN project \citep[][]{kanna20201_thesan_intro,garaldi2021_thesan_gal_IGM} has recently performed a series of AREPO-RT \citep[][]{kannan_arepo-rt_2019} simulations of the EoR. While they reach high spatial resolution inside of galaxies, \codaiii \,  improves on their approach in at least 2 aspects: their fiducial run, THESAN-1, stops at z=5.5, whereas we were able to go beyond this redshift, down to z=4.6, and as a result compare \codaiii \, \mfpacro \, with observations in a larger redshift range, which, most importantly, includes post-reionization measurements as well. Also, THESAN-1 uses a typical speed of light reduction of 0.2, which can have a strong impact on the post-reionization residual neutral hydrogen fraction \citep[][]{ocvirk_impact_2019}, and to some extent change the timing of reionization \citep[][]{deparis_impact_2019}, complicating the interpretation of their simulation results. To avoid such difficulties, Cosmic Dawn III uses the full speed of light thanks to the GPU-optimized radiative transfer module CUDATON \citepalias[][]{ocvirk_cosmic_2016,ocvirk_cosmic_2020}.

First, we present our methodology, including the particularities of our simulation code RAMSES-CUDATION, the simulation setup, and the computational steps we take to derive the \mfpacro. Then, we present the global properties of reionization in the \codaiii \, simulation, and the evolution of the \mfpacro \, in \codaiii. Finally we conclude our work.

\section{Methods}

\subsection{Simulation}

\subsubsection*{Overview of RAMSES-CUDATON}

The \codaiii \, (Cosmic Dawn III) simulation was performed using the fully coupled radiation and hydrodynamics code RAMSES-CUDATON \citepalias[presented in][]{ocvirk_cosmic_2016,ocvirk_cosmic_2020}. Here we briefly recall its main properties as well as details concerning its deployment.

RAMSES-CUDATON results from the coupling of the RAMSES code for N-body dark matter dynamics, hydrodynamics, star formation and stellar feedback \citep[][]{teyssier_cosmological_2002}, with  the radiative transfer module ATON \citep[][]{aubert_radiative_2008}, which handles the radiative transfer of ionising photons via the M1 method \citep[][]{levermore_relating_1984}, hydrogen photo-ionisation and thermo-chemistry. Motivated by the important computational cost of radiative transfer, we accelerate the ATON module using GPUs \citep{aubert_reionization_2010}, whilst the remaining RAMSES physics run on CPUs. The performance boost thus obtained allows us to run fully coupled simulations with a full speed of light, circumventing pitfalls encountered by certain reduced speed of light approaches \citep{gnedin_proper_2016,deparis_impact_2019}, in particular on the residual neutral fraction after overlap \citep{ocvirk_impact_2019}. 

\subsubsection*{Improvements over our previous simulation \codaii \,}
For the most part, the code and simulation setup are as described in \citetalias{ocvirk_cosmic_2020}, but with higher resolution. Other features of \codaiii \, which differ from past CoDa simulations were fully tested and calibrated by a large suite of smaller simulations of the same numerical resolution as \codaiii \, to ensure a good match to observational constraints on the EoR. While the chief purpose of these calibration runs was to prepare for \codaiii, the calibration runs themselves were novel enough to warrant new science as described in \cite{ocvirk_lyman-alpha_2021}. Since the new features of \codaiii \, were already described there, we refer the reader to that paper for more detail.
We quickly enumerate the novelties of \codaiii \, with respect to \codaii \, hereafter. 
\begin{itemize}
\item \codaiii \, uses a $8192^3$ grid while \codaii \, had a $4096^3$ grid, for the same volume, meaning its physical resolution is a factor of two finer, and its mass resolution is eight times higher.

\item We take a more detailed approach with respect to our source model than in the previous CoDa simulations. In \codaiii, each stellar particle receives an ionising emissivity value based on its mass, age, and metallicity. We pre-compute emissivities using BPASS V2.2.1 \cite{eldridge_population_2020}, considering that each stellar particle is a star cluster of about \SI{e4}{\simsun} forming over a few Myr as in \citetalias{ocvirk_lyman-alpha_2021}.
As a result, in \codaiii \, individual stellar particles' ionising luminosities decrease more gradually and realistically than in \codaii{. \footnote{{In the previous CoDa simulations, stellar particles had a fixed stellar luminosity until their massive stars went supernova, at which point their ionising luminosities were set to 0.}} }

\item One of the main model differences with respect to \codaii \, lies within the star formation sub-grid model. In \codaiii, for a gas cell to form stars, we not only require that it be dense enough, but also that it be cooler than \SI{2e4}{\kelvin}. This additional temperature threshold for star formation has been widely used in similar simulations before \codaiii \, including \codai \, \citep[][]{ocvirk_cosmic_2016} and others \citep[e.g.][]{dubois_onset_2008}. Its physical interpretation is as follows: gas cells hotter than the threshold are so because of shocks, usually caused by supernovae, which are not favoured sites for star formation until they manage to cool back down to below the \SI{2e4}{\kelvin} threshold.

We also found that including the temperature threshold allowed our simulations to achieve better agreement with the residual neutral fraction at the end of the EoR and with the ionisation rate of ionized regions, thus advocating its use in \codaiii. For more information and a comparison between star formation with, and without the temperature threshold, we refer the reader to \citetalias{ocvirk_lyman-alpha_2021}. 

\item We use a sub-grid stellar escape fraction of $\rm f_{esc}^{sub}=1$, whereas we used $\rm f_{esc}^{sub}=0.42$ in \codaii, i.e. all photons produced by the stellar particles in \codaiii \, are released in the host cell. {We shall fully investigate the escape of ionising photons in an upcoming paper, however we can already tentatively explain why we require a higher $\rm f_{esc}^{sub}$ in \codaiii{}: First, and due to the changes to the star formation model, star formation in \codaiii{} is not permitted in very hot ionised cells. For the same amount of star formation it is then simple to imagine that one requires more photons to complete Reionization in \codaiii{} with respect to \codaii{}. Second, \codaiii{}'s higher resolution likely gives rises to more and denser clumps of gas in the IGM, potentially further raising the required number of photons to complete Reionization. Third, the ionising emissivity of stellar particles evolves in a smoother fashion in \codaiii{}, plausibly altering the balance between ionization and recombination (although we found this effect to be relatively moderate in smaller test volumes).
}

\item Just as in \codai \, \& \codaii \, (the two previous Cosmic Dawn simulation projects), upon reaching \SI{10}{\mega\year} of age, a mass fraction $\eta_{\rm SN}=0.2$ of stellar particles detonates as supernovae. Each supernova event injects \SI{51}{\ergs} of energy for every \SI{10}{\simsun} of progenitor mass into its host cell \citep[using the kinetic feedback of][]{dubois_onset_2008}. After the supernova event, a long lived particle of mass $\rm (1-\eta_{SN})M_{birth}$ remains. Unlike \codai \, and \codaii, \codaiii \, features the standard RAMSES chemical enrichment\footnote{for clarification: this is not a new code feature, but was not activated at run time in the previous simulations, for simplicity.}. Metals are produced by supernovae. A fraction $\rm y=0.05$ of the total supernova mass is re-injected as metals. Metallicity is treated as a passive scalar, and metals are thus advected along with the gas.

\item \codaiii \, features a physically motivated model for the formation and destruction of dust in galaxies on the fly, that we coupled to the ionizing radiative transfer of ATON \citep[see][for details]{lewis_dustier_2022}. An independent paper on the dust model is also in preparation (Dubois et al.). The dust aspects of \codaiii \, will be analysed in a forthcoming paper, and can be left aside for the present study.
\end{itemize}


For quick reference, Table \ref{tab:codaiii} recaps the essential physical and numerical parameters of \codaiii.

\subsubsection*{Initial conditions \& Cosmology}

As in previous CoDa simulations, \codaiii \,'s initial conditions were constrained so as to produce facsimiles of the progenitors of the Local Group \citep[See][]{sorce_cosmicflows_2016, sorce_galaxy_2018} within the Constrained Local Universe Simulations project (CLUES\footnote{https://www.clues-project.org/cms/}). We will use this unique opportunity to investigate the reionization of the Local Group in a forthcoming paper. For the present study and for the purpose of this paper, we can forgo the Local Universe aspect and treat \codaiii \, as a random realization of the Universe, as its mass function, for instance, is compatible with that of a random patch of Universe. The cosmology used ($\Omega_\Lambda=0.693$, $\rm \Omega_m=0.307$, $\rm \Omega_b=0.0482$, $\rm H_0=$ \SI{67.77}{\kilo\meter\per\second\per\mega\parsec}, $\rm \sigma_s=0.829$, $\rm n=0.963$) here is taken from the Planck 2014 constraints \citepalias[][]{collaboration_planck_2014}\footnote{these constraints are compatible with subsequent Planck publications}, as summarised in Table \ref{tab:codaiii}. The initial conditions were generated at $\rm z=150$.

\subsubsection*{Deployment}

\codaiii \, was run on Summit at the Oak Ridge National Laboratory / Oak Ridge Leadership Computing Facility. It was deployed on 131072 processors coupled with 24576 GPUs, distributed on 4096 nodes, and ran for $\sim$ 10 days, allowing \codaiii \, to reach a final redshift of z=4.6, i.e. sufficiently late after overlap that we can consider reionization complete and sample well into the post-overlap phase. Snapshots were written every $\sim 10$ Myr, we use a subset of them to derive the properties of the IGM in the following section. Subsequent processing {of the more than 20 PB of data outputs} took place at the same facility on Andes and Summit.

\begin{table}
\centering
\begin{tabular}{l l}
\hline
\multicolumn{2}{c}{Setup} \\
\hline
 
Grid size \& Dark matter particle number  & $8192^3$\\
Box size & \SI{94.43}{\comovmega\parsec} \\
Force resolution & \\
\hspace{0.3cm} ...comoving & \SI{11.53}{\comovkilo\parsec} \\
\hspace{0.3cm} ...physical ($\rm z=6$) & \SI{1.65}{\kilo\parsec}\\

Dark matter particle mass & \SI{5.09e4}{\simsun}\\

Average cell gas mass & \SI{9.375e3}{\simsun}\\

Stellar particle mass & \SI{11732}{\simsun} \\

\hline
\multicolumn{2}{c}{Star formation \& feedback} \\
\hline
 
Density threshold for star formation & 50<$\rm \rho_{gas}$>\\
Temperature threshold for star formation & \SI{2e4}{\kelvin}\\
Star formation efficiency $\rm \epsilon_\star$ & 0.03\\
Massive star lifetime & \SI{10}{\mega\year}\\
Supernova energy & \SI{e51}{\ergs}\\ 
Supernova mass fraction, $\rm \eta_{SN}$ & 0.2 \\
Supernova ejecta metal mass fraction  & 0.05\\

\hline
\multicolumn{2}{c}{Radiation} \\
\hline

Stellar ionising emissivity model & BPASS V2.2.1 binary \\
{\scriptsize  \citep[from][]{eldridge_population_2020}} & \\
Stellar particle sub-grid escape fraction, $\rm f_{esc}^{sub}$ & $1.0$ \\

Effective photon energy & \SI{20.28}{\ev}\\
Effective HI cross-section (at \SI{20.28}{\ev}) & \SI{2.493e-22}{\meter\squared}\\

\hline
\multicolumn{2}{c}{Reionization} \\
\hline

CMB electron scattering optical depth, $\rm \tau_{es}(z=14)$ & 0.0497 \\
reionization mid-point, $\rm z_{re,\,50\%}$ & 6.81 \\
reionization complete, $\rm z_{re,\,99.99\%}$ & 5.53 \\

\hline

\end{tabular}
\caption{Overview of the \codaiii \, simulation.}
\label{tab:codaiii}
\end{table}

\subsection{Computing the mean free path of ionising photons}
\label{sec:mthds}

To compute the \mfpacro, we employ four approaches: 
\begin{itemize}
    
\item First, we employ a method (called average transmission method from hereon) inspired by \cite{kulkarni_models_2016} . For a given simulation snapshot, we draw 1024 lines of sight (LoS) across the whole simulation box in one direction. Our LoS are evenly separated so as to evenly sample the box.

For each LoS, we then compute the transmitted flux at \SI{912}{\angs} as a function of position along the LoS x. These can then be averaged into one average transmitted flux F(x), from which we obtain \mfp \ by fitting it with Eq. \ref{eq:mfp} for the parameter $\lambda$:

\begin{equation}
\begin{array}{lll}
    \rm F(x) & = & \rm F_0 exp\Big(-\frac{x}{\lambda}\Big)
    \label{eq:mfp}
\end{array}
\end{equation}

where $\rm F_0$ is a normalisation parameter, equivalent to a flux incident where $\rm x=0$, that is free during the fit.
For key redshifts, we check that our results are converged with respect to the number of lines of sight, and find no significant difference in our result with 8x or 64x more lines of sight.
In order to mimic the small number statistics regime of the observations, we draw samples of only 13 LoS \citepalias[the same LoS sample size as in][at z=6]{becker_mean_2021} from our full sample of 1024. We repeat this 10$^5$ times, and our final \mfp \, result is then the median of these $10^5$ realizations. Also of interest, this approach yields the standard deviation of this set of mfp.

\item Second, in order to access the distribution of \mfpacro \, belonging to individual lines of sight (\mfplos \, hereafter), we fit Eq. \ref{eq:mfp} to the transmitted flux of every LoS individually. We can then also compute the mean, median, and spread of the resulting distribution at every redshift. The results from bad fits (reduced $\rm \chi^2 > 1.0$) are discarded. {This occurs only 5 times for the $1024\times14$ LoS that we process, and correspond to lines of sight that cross a large overdensity. In Appendix \ref{app:check}, we compare this method to that of \cite{rahmati_mean_2018} \citep[also found in][]{garaldi2021_thesan_gal_IGM}, and show that both methods for computing the mean free path along individual LoS are remarkably consistent.}

\item Third, as we will see in Sec. \ref{sec:mfp}, our \mfplos \, are divided into two families that reside in neutral, and ionized areas of the IGM. We can easily isolate the evolution of the \mfplos \, from the ionized regions so as to more closely mimic the observational context of the measurement of \mfpacro, since such measurements cannot be performed in still neutral medium. To this end, for every redshift we divide the distribution of \mfplos \, in two modes, at the approximate position of the minimum of the distributions and therefore the transition between the modes, and consider only the long mode of the distribution, corresponding to ionized LoSs. We refer to this approach as the ionised method. We use the notation \mfpion \, to refer to mfps computed in this manner.

\end{itemize}

\section{Results}

\subsection{Global properties of the IGM during the EoR}
 
\begin{figure*}
    \centering
    \includegraphics[width=0.49\textwidth]{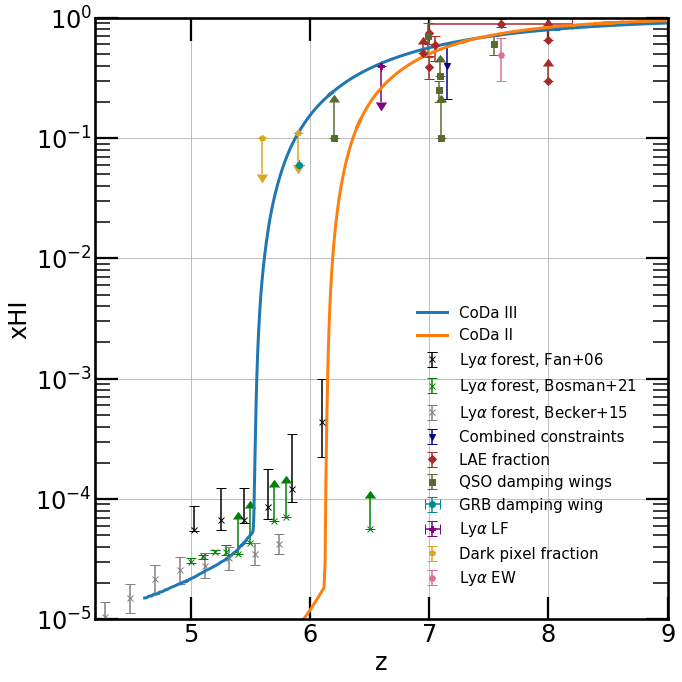}
    \includegraphics[width=0.49\textwidth]{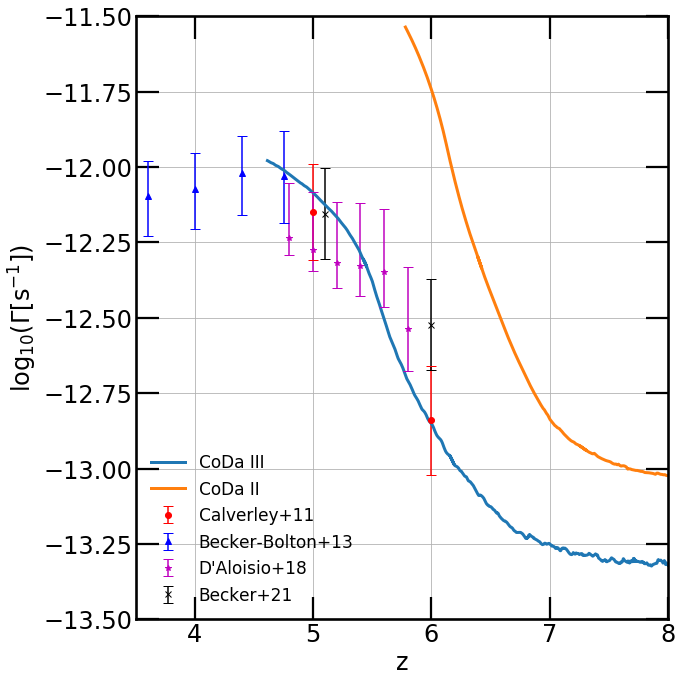}\\
    \includegraphics[width=0.49\textwidth]{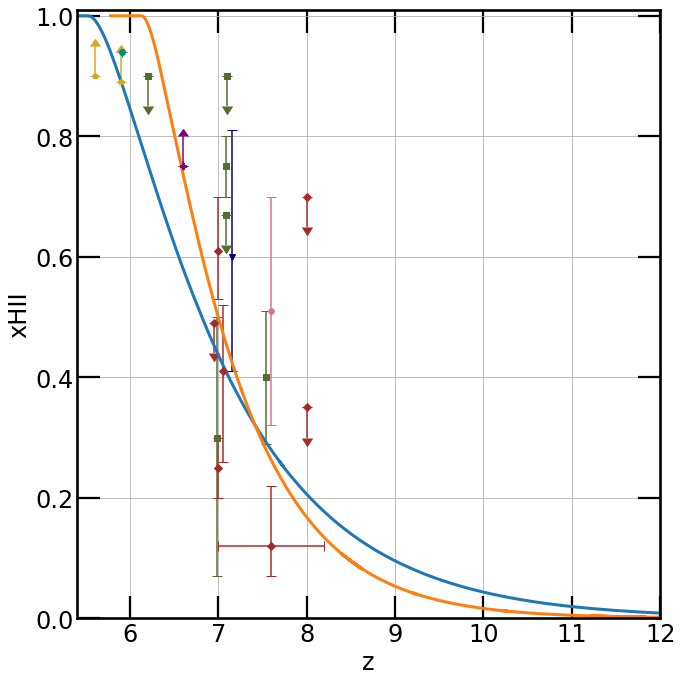}    
    \includegraphics[width=0.499\textwidth]{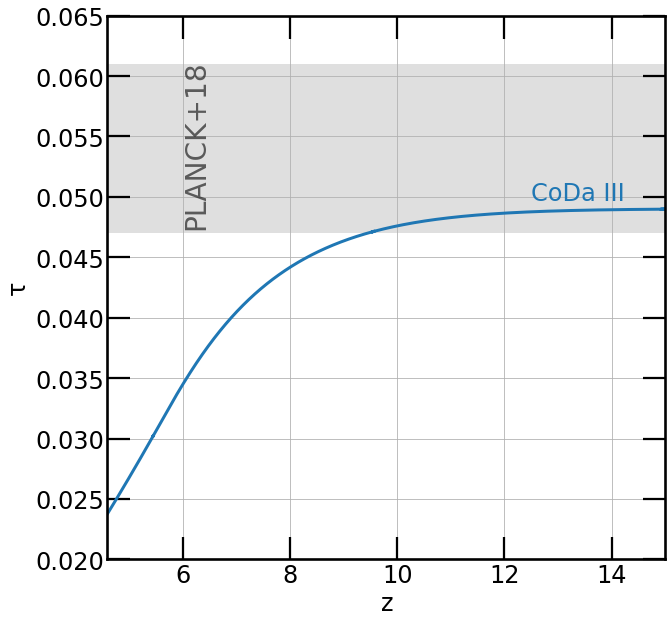}
    
    \caption{\emph{Top left}: Volume weighted average neutral fraction along with a collection of constraints including \protect \cite{fan_constraining_2006,becker_evidence_2015,bosman_hydrogen_2022,ouchi_statistics_2010,mcgreer_model-independent_2015,totani_high-precision_2016,wang_low-redshift_2021,durovcikova_reionization_2020,banados_800-million-solar-mass_2018,schroeder_evidence_2013, mortlock_luminous_2011, tilvi_rapid_2014, pentericci_new_2014,schenker_line-emitting_2014, ono_spectroscopic_2012, mason_universe_2018,hoag_constraining_2019, greig_xHI_2017,jung_galaxies_2021}. 
    \emph{Top right}: Ionisation rate in ionized regions when compared to constraints from \protect \cite{calverley_measurements_2011, becker_new_2013, daloisio_large_2018, becker_mean_2021}. 
    \emph{Bottom Left}: Volume weighted ionised fraction. 
    \emph{Bottom Right}: Optical depth to CMB photons due to scattering by free electrons, compared to the Planck collaboration results \protect \citetalias{planck_collaboration_2020}. {\codaiii \, is in very good agreement with constraints on the history of ionisation of the IGM as well as on the ionization rate, whilst remaining compatible with the optical depth due to electron scattering measured by Planck, despite a rather late end of reionization. Overall the improvement over \codaii  \, is very significant.}}
    \label{fig:eor}
\end{figure*}

The top left panel of Fig. \ref{fig:eor} shows the evolution of the average volume weighted neutral fraction of hydrogen (\xhi) in \codaiii \, and \codaii, along with a range of constraints. In both simulations, \xhi \, starts to drop noticeably between $\rm z=10$ and $\rm z=9$. However, in \codaiii, reionization ends later and less abruptly than in \codaii, only finishing by $\rm z \approx 5.6$ ($\rm \approx 6$ in \codaii). Further, in \codaii \, the final average value of the neutral fraction is far lower than in \codaiii. This latter difference is due to the combination of the new star formation sub-grid model and improved resolution, as explained in \citetalias{ocvirk_lyman-alpha_2021}. Considering this, along with the rather late reionization obtained \citep[compared to the canonical value of z=6 for the end of reionization in use since][]{fan_constraining_2006}, \codaiii \, is in much better agreement with the constraints from the transmission in the \lya \, forest (and modelling) on \xhi \, after reionization. In particular, the agreement with the data from \citetalias{bosman_hydrogen_2022} is excellent. The match with constraints from \cite{becker_evidence_2015} is also remarkable in \codaiii \, after the ankle of the \xhi \, curve, although   \cite{becker_evidence_2015} favours slightly earlier reionization than in \codaiii. Finally we find that
 the older constraints from \cite{fan_constraining_2006} are disfavoured both by the more recent determinations of \citetalias{bosman_hydrogen_2022} and \cite{becker_evidence_2015}, and our own simulations.
 The bottom left panel of Fig. \ref{fig:eor} shows the evolution of the average ionised fraction of Hydrogen ($\rm x_{HII}$) on a linear scale, along with the same constraints. \codaiii \, is in better agreement with a greater fraction of the collection of constraints for $\rm x_{HII} \gtrsim 0.1$ than \codaii. 
 Overall the agreement of \codaiii \, with modern constraints on the evolution of the progress of reionization\, is very good.

The top right panel of Fig. \ref{fig:eor} shows the evolution of the ionisation rate in ionized regions (\gammahi) \, in both \codaiii \, and \codaii. In the former, \gammahi \, rises rapidly by roughly a factor of 10 between $\rm z=7$ and $\rm z=5.5$, after which the slope becomes shallower. \codaiii \,'s \gammahi \, is a very good fit to the mix of observational constraints available, and a great improvement on the \gammahi \, from \codaii, which overshoots all constraints for the presented redshift range. This excess is symptomatic of \codaii \,'s excessive ionisation of the IGM \citepalias[see][for discussions about this]{ocvirk_cosmic_2020,ocvirk_lyman-alpha_2021}. As with \xhi, the origin of this improvement lies both in the switch to a late reionization calibration and in the improved star formation sub-grid model, as well as increased resolution \citepalias[as discussed in][]{ocvirk_lyman-alpha_2021}.

The bottom right panel of Fig. \ref{fig:eor} shows that although reionization ends later in \codaiii, the optical depth due to scattering on free electrons seen by the Cosmic Microwave Background remains compatible with the constraints given by the latest Planck results \citepalias[][]{planck_collaboration_2020}, demonstrating that a late reionization scenario may still yield a high enough electron scattering optical depth\footnote{An even higher optical depth may still be achieved by adding the contribution from mini-halos at $\rm z \geq 15$, adding an extra $\rm \Delta \tau_{\rm z\geq 15} \approx 0.008$, that are not treated in this work but suggested by \cite{ahn_cosmic_2021}.}.

\begin{figure}
    \centering
    \includegraphics[width=0.45\textwidth]{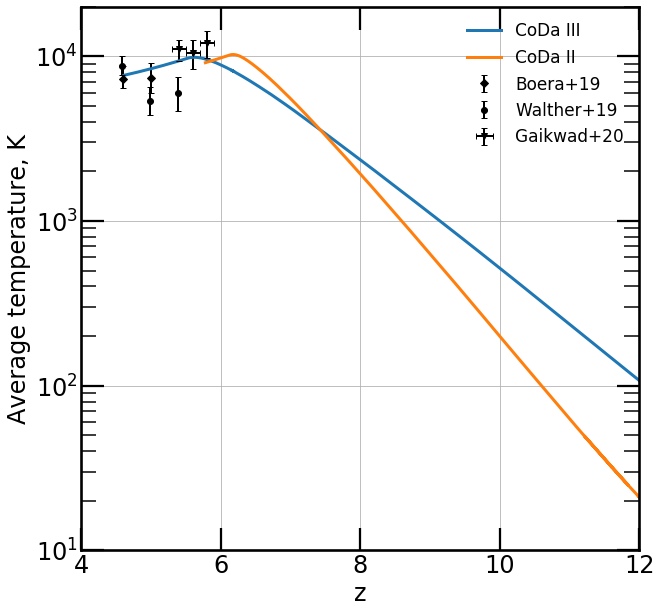}
    \caption{Average evolution of the gas temperature in \codaiii{} (blue) and \codaii{} (orange). For comparison, constraints from \protect \cite{boera_revealing_2019, walther_new_2019, gaikwad_probing_2020} are also shown.}
    \label{fig:temps}
\end{figure}

{Fig. \ref{fig:temps} shows the average evolution of the gas temperature in \codaiii{} (blue) and \codaii{} (orange). In both simulations, the average temperature rises as Reionization progresses and a larger and larger fraction of the IGM is photo-heated. Eventually, the temperatures reach a characteristic peak as Hydrogen Reionization completes, before cooling slightly. We find that the average temperature in \codaiii{} (and to some extent in \codaii{}) is consistent with recent constraints when available ($\rm 6\geq z \geq4.6$). \codaii{} heats up much more rapidly, which is coherent with the earlier Reionization and higher photo-ionisation rates shown in Fig. \ref{fig:eor}.}

Overall reionization in \codaiii \, reaches a better agreement with observational constraints on the IGM and other works from the literature than \codaii \, did, signalling a marked improvement in our description of reionization.

\subsection{Results: the mean free path of ionising photons}

\label{sec:mfp}

\begin{figure}
    \centering
    \includegraphics[width=0.48\textwidth]{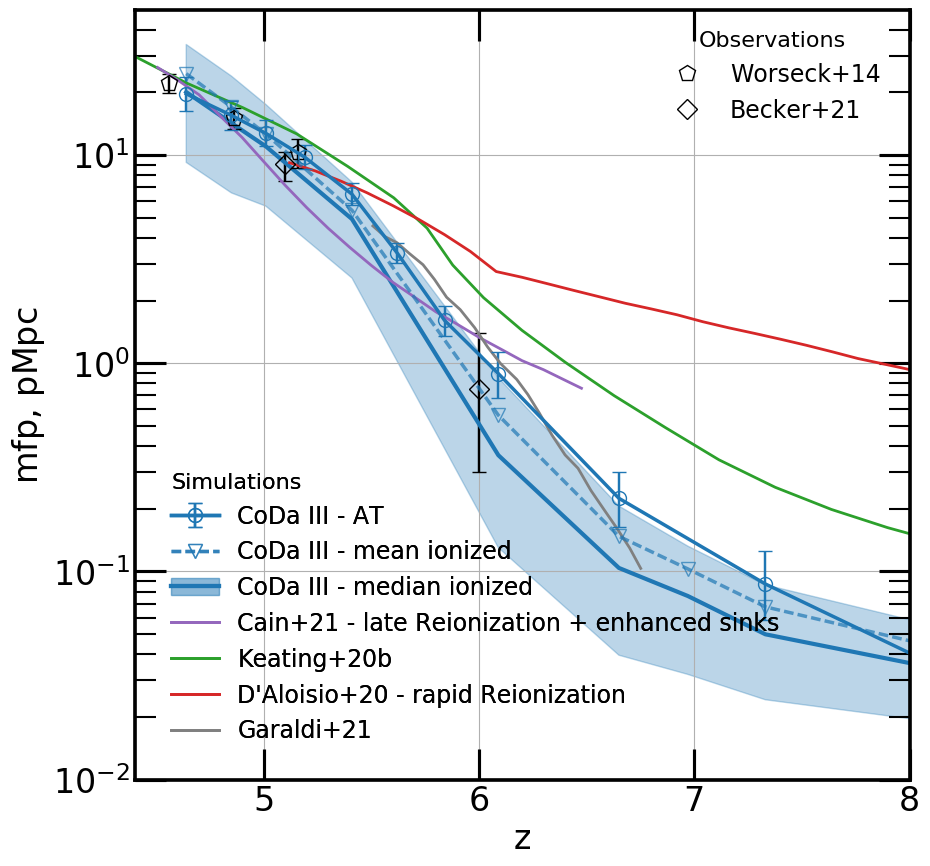}
    \caption{Evolution of the mean free path with redshift in \codaiii, and comparison to observational constraints from \protect \citep[][]{worseck_giant_2014, becker_mean_2021} and previous simulation results from the literature \protect \citep[][]{keating_long_2020, daloisio_hydrodynamic_2020, cain_short_2021, garaldi2021_thesan_gal_IGM}. The blue graphics show results from \codaiii. The full blue line with circles and errorbars show the median and $\rm 25^{th}-75^{th}$ percentile region for \mfp. The full blue line with triangles shows the mean \mfpion, the blue dashed line and blue region show the median and $\rm 25^{th}-75^{th}$ percentile region \mfpion. The median mean free path in \codaiii \, is in good agreement with existing observational constraints from $\rm z=4.5$ till $\rm z=6$.}
    \label{fig:mfp}
\end{figure}

We now move to our titular subject: the mean free path of ionising photons. Fig \ref{fig:mfp} shows the evolution of \mfpacro \, over time in \codaiii. Observations from \cite{worseck_giant_2014} and \citetalias{becker_mean_2021} show that the \mfpacro \, increases over time for $\rm z<6$. The full blue line with errorbars shows that in \codaiii, \mfp \, increases from $\approx \, $\SI{1}{\propmega\parsec} at  $\rm z \approx 6$ to $\approx \,$\SI{20}{\propmega\parsec} at $\rm z\approx 4.6$. In fact, \mfp \, grows faster between $\rm z\approx 5.75$ and $\rm z\approx 5.2$, allowing \codaiii \, to match the newest low mfp from observations at $\rm z = 6$ \citep{becker_mean_2021} and the high observed mfp at $\rm z \lesssim 5.25$, whereas most other theoretical work has tended to overshoot the observed mfp \, at z=6. Because of the stronger evolution of the mfp \, in \codaiii, we report values that are significantly lower than in other theoretical work for $\rm z \gtrsim 6$, where observations are very challenging. This points to substantially different progressions of reionization in the IGM of these other simulations, feasibly implying differences in ionising emissivities and source populations.

Already, previous work \citepalias{davies_predicament_2021,cain_short_2021} has highlighted that the latest determinations of mfp by \citetalias{becker_mean_2021} point to a late and rapid reionization,  which \codaiii \, matches.
However, unlike in \cite{cain_short_2021} we do not need to invoke unresolved photon sinks in order to obtain our agreement with the observed mfp.

The THESAN simulation \citep{garaldi2021_thesan_gal_IGM} also finds a good agreement with the $\rm z=6$ data point from \citetalias{becker_mean_2021}, however, unfortunately, the simulation ended too soon to comment on their agreement with lower redshift constraints, and it is therefore difficult to gauge the overall success of their simulation in reproducing the evolution of the properties of the IGM over the whole EoR, down to the post-overlap regime. In any case, it is possible that fully coupled RHD simulations are better able to match the mfp constraint from \citetalias{becker_mean_2021} at $\rm z=6$ than post-processing or modelling methods, perhaps because of the more consistent treatment of the radiative-hydrodynamical interactions in the IGM and possible radiative feedback in galaxies, and/or the superior resolution.

\begin{figure}
    \centering
    \includegraphics[width=0.48\textwidth]{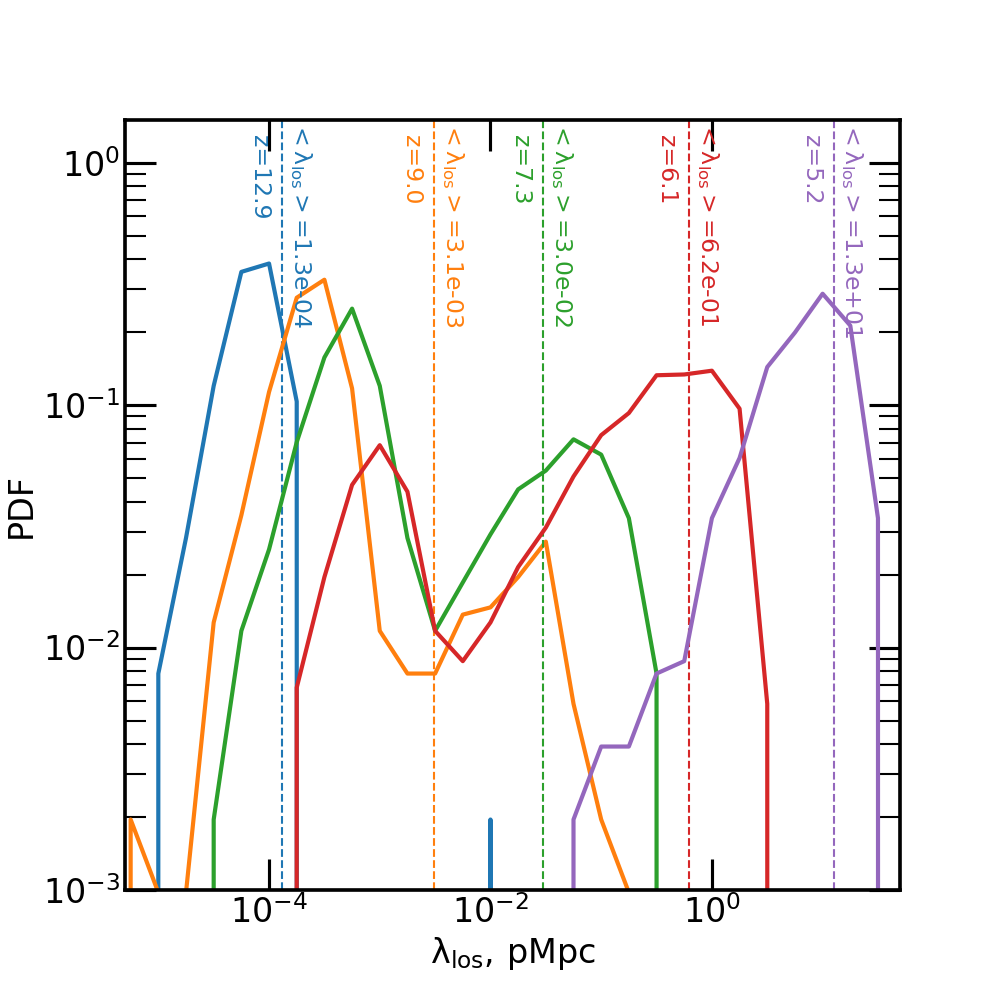}
    \caption{Distribution of \mfplos. Dashed vertical lines show the distribution mean for each redshift. For readability, only a subset of redshifts roughly 200 Myr apart between z=12.9 and z=5.2 is shown. {The \mfplos \, are distributed in a neutral opaque mode of short mean free paths, and an ionised transparent mode of long mean free paths. The relative weight of these modes changes as reionization progresses.}}
    \label{fig:mfp_distrib}
\end{figure}

In order to gain further insight into what drives the mfp measured in \codaiii, we show the distribution of individual LoS \mfpacro \, (i.e. \mfplos)\, at several redshifts in  Fig. \ref{fig:mfp_distrib}. At $\rm z=12.9$, before reionization is fully underway, the distribution of \mfplos\, has a single mode near \SI{e-4}{\propmega\parsec}. Although this is shorter than the simulation's cell size, it is physical: indeed, the general formula for the mean free path of ionising photons reads $1/\sigma \rho_{\rm HI}$, where $\sigma$ is the interaction cross-section and $\rho_{\rm HI}$ the number density of neutral hydrogen atoms. In the dense, neutral high-redshift Universe before reionization, mean free paths shorter than our cell size naturally occur and are fully consistent with the measurements presented here. 

By $\rm z=7.3$, reionization is in full swing, and the distribution of \mfplos \, is now bi-modal, displaying a short and a long mode. The short mode peaks around \SI{8e-4}{\propmega\parsec} and seems to descend from the $\rm z=12.9$ mode. However, approximately 30\% of LoS are now scattered around a secondary, longer mode peaking near \mfplos=\SI{e-1}{\propmega\parsec}. This bi-modality reflects the co-existence of large fully ionized regions already by $\rm z=7.3$, when reionization is almost complete in half of \codaiii \,'s volume (about 40\% ionized reading from the bottom left panel of Fig. \ref{fig:eor}). At $\rm z=6.1$, both modes have moved to slightly higher \mfplos \, values (\SI{e-3}{\propmega\parsec} for the short \mfplos \, peak, and \SI{0.64}{\propmega\parsec} for the long \mfplos \, peak), and the relative weight between modes has swapped: now most LoS belong to the long \mfplos \, peak. 

By $\rm z=5.2$, reionization is complete, all of the LoS have been ionized and the short mode has disappeared in favour of the long mode. The evolution of the \mfplos \, distribution reveals that the \codaiii \, \mfplos \, evolves due to both gradual changes along every LoS (due to cosmic expansion driving the average physical density to decrease over time), and to rapid changes along short \mfplos \, LoS due to reionization. The rapid increase of \mfp \, in Fig. \ref{fig:mfp} between $\rm z=6.1$ and $\rm z=5.2$ could be driven by the disappearance of the shortest \mfplos \, values as more and more LoS experience reionization.

Fig. \ref{fig:mfp_distrib} shows that the transition between the ionised mode of the \mfplos \, distribution and the neutral mode occurs near \SI{e-2}{\propmega\parsec}). Accordingly we select only those \mfplos \, longer than this threshold for the ionised method. The blue solid curve with triangles in Fig. \ref{fig:mfp} shows the evolution of  \mfpion \, in \codaiii. For $\rm z>5$, both the mean and the median \mfpion \, are lower than \mfp. This makes our prediction based on the mean \mfpion \, a better match to the $\rm z=6$ constraint from \citetalias{becker_mean_2021}, and also shows that the average transmission method (measuring the mean free path from an average simulated transmission spectrum) is biased towards high transmissions and overestimates the median \mfplos \, at the redshifts considered in Fig. \ref{fig:mfp}.
For $\rm z\leq5$, the mean \mfpion \, is slightly above the constraints from \cite{worseck_giant_2014}, however the median remains in very good agreement and there is a broad scatter, comparable to the error bar of \citetalias{becker_mean_2021}.

\section{Conclusions}

In this article, we have presented first results from our latest large scale simulation of the Epoch of Reionization, Cosmic Dawn III (\codaiii \,){, the largest fully coupled RHD simulation of Reionization ever performed.} It improves on our previous \codaii \, simulation in many aspects, such as using a two times higher spatial resolution and 8 times higher mass resolution, as well as updates of the physical modules and sub-grid models, in particular star formation and stellar source modelling. We show that, {thanks to these improvements, }\codaiii \, compares much better than \codaii \, with observational constraints of the high-redshift Universe, in particular the evolution of the neutral hydrogen fraction which sees a dramatic improvement, as well as the ionizing rate in ionized regions, while the evolution of the ionized fraction and the electron scattering optical depth are also in good agreement with the constraints available. 
Moreover, we showed that \codaiii \, is the first fully coupled radiative hydrodynamics simulation to reproduce the latest constraints on the ionising photon mean free path all the way from $\rm z=6$ down to $\rm z=4.6$ thanks to a late reionization ($\rm z_{rei}=5.6$) and a rapid evolution of the ionising mean free path. 
In particular, \codaiii \, shows that the surprisingly low mean free path measured by \citetalias{becker_mean_2021} can naturally occur in a well-calibrated simulation of the EoR such as ours.
The distribution of mean free paths during reionization is generally bi-modal, displaying a short, neutral mode and a long, ionized mode. {Both modes shift to higher \mfpacro \, as reionization progresses: The neutral mode peak shifts from sub-kpc to kpc scales of \mfpacro, and the ionized mode peak from 0.1 Mpc/h (at z=7) to $\sim 10$ Mpc/h at ($\rm z=5.2$).} The balance of these modes reflects the degree of ionisation of the simulated volume. The bi-modal distribution reduces to a single short (long) mode before (after) reionization, respectively. 
Finally, by using 3 different methods for computing the mean free path from \codaiii, we show that the choice of the method is important. In particular, measuring the mean free path from an average transmitted flux stacking a number of lines of sight tends to overestimate the mean free path otherwise obtained as the average or the median of the mfp of individual lines of sight.
Overall, \codaiii's description of reionization in the IGM appears to be a marked improvement over \codaii, providing an adequate basis with which to perform a broad range of studies of the EoR, such as \lya \, transmission in the IGM, but also of star formation suppression, and the ionising photons budget of galaxies.

\section*{Acknowledgements}
The authors thank F. Davies and H. Katz for fruitful discussions. Cosmic Dawn III was performed on Summit at Oak Ridge National Laboratory / Oak Ridge Leadership Computing Facility (Project AST031) using an INCITE 2020 allocation. JL acknowledges support from the DFG via the Heidelberg Cluster of Excellence STRUCTURES in the framework of Germany’s Excellence Strategy (grant EXC-2181/1 - 390900948). JS thanks Sergey Pilipenko for sharing the latest version of the ginnungagap code as well as for his guidance with using it. JS acknowledges support from the ANR LOCALIZATION project, grant ANR-21-CE31-0019 of the French Agence Nationale de la Recherche. JS gratefully acknowledges the Gauss Centre for Supercomputing e.V. (www.gauss-centre.eu) for providing computing time to build the initial conditions of the simulations for this project on the GCS Supercomputer SuperMUC-NG at Leibniz Supercomputing Centre (www.lrz.de). GY acknowledges financial support from MICIU/FEDER under project grant PGC2018-094975-C21. KA was supported by NRF-2016R1D1A1B04935414, 2021R1A2C1095136, and 2016R1A5A1013277. ITI was supported by the Science and Technology Facilities Council [grant numbers ST/I000976/1 and ST/T000473/1] and the Southeast Physics Network (SEP-Net). HP was supported by the World Premier International Research Center Initiative (WPI), MEXT, Japan and JSPS KAKENHI grant No. 19K23455.

\section*{Data availability}

The authors will share the data used for this article upon reasonable request.


\bibliographystyle{mnras}
\bibliography{lettre_mfp} 



\appendix{}

\section{Consistency check of mean free paths for individual lines of sight}

\label{app:check}

\begin{figure*}
    \centering
    \includegraphics[width=0.48\textwidth]{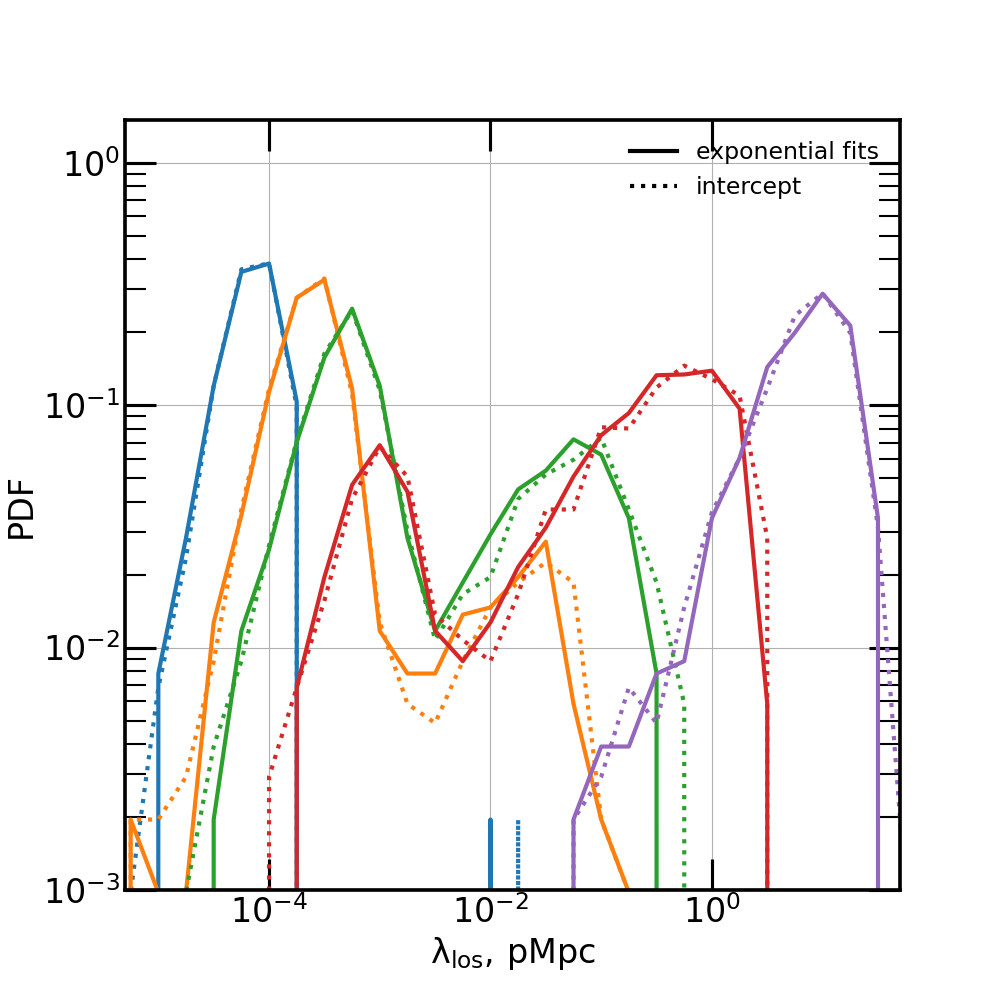}
    \includegraphics[width=0.48\textwidth]{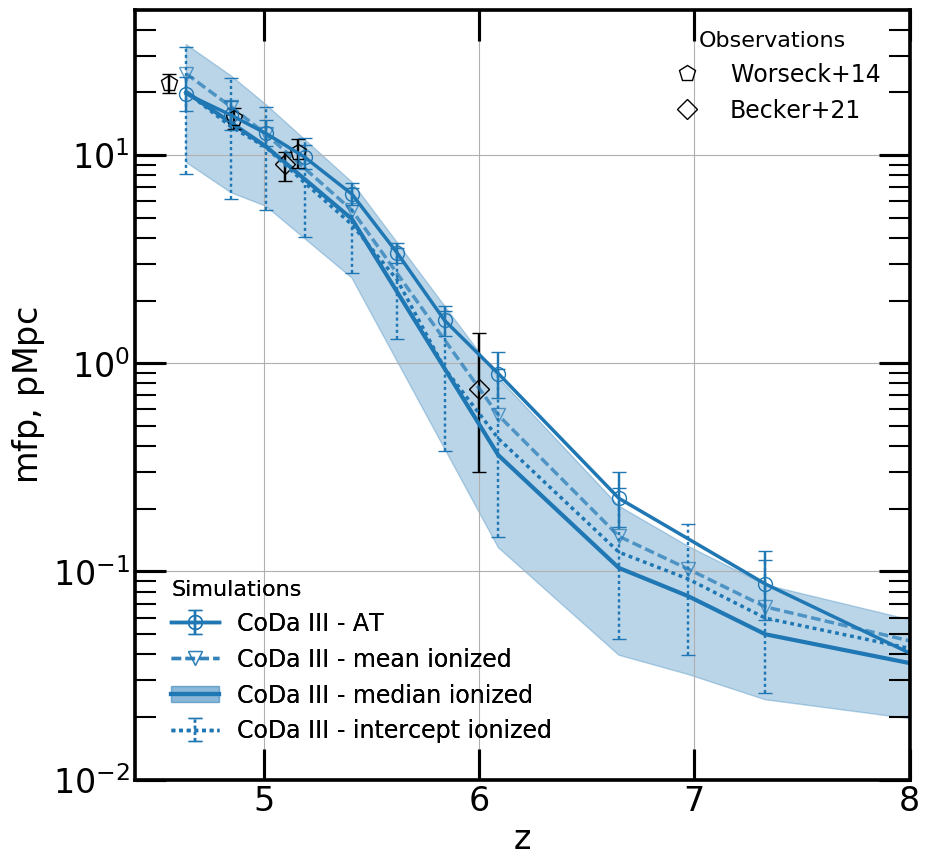}
    \caption{\emph{Left: } Distribution of \mfplos. Dashed vertical lines show the distribution mean for each redshift. For readability, only a subset of redshifts roughly 200 Myr apart between z=12.9 and z=5.2 is shown. The \mfplos \, are distributed in a neutral opaque mode of short mean free paths, and an ionised transparent mode of long mean free paths. The relative weight of these modes changes as reionization progresses.
    \emph{Right: } We compute the evolution of the mean free path along ionised lines of sight, by applying the same methodology to the intercept distribution as used to compute the \mfplos{}. The dotted lines show the result when compared to the other techniques we employ.}
    \label{fig:mfp_distrib_rahmati}
\end{figure*} 

{In order to verify the \mfplos{} and \mfpion{} values for individual lines of sight, we also compute these using another method found in the literature \citep[e.g.][]{rahmati_mean_2018}. Using Eq. \ref{eq:mfp}, one finds that:

\begin{equation}
\begin{array}{l}
    \rm \frac{F(x)}{F_0}  = exp(-1), \, if \, x = \lambda
    \label{eq:mfp_intercept}
\end{array}
\end{equation}

Applying this to each LoS, we search for the cells whose transmission $\rm T_a(x_a)=F_a(x_a)/{F_0}$ and $\rm T_b(x_b)$ bracket $\rm exp(-1)$. We then interpolate between the cells to find x so that $\rm T(x)=exp(-1)$. There are two limit cases that we must deal with. First, we must treat the case of very transparent LoS for which $\lambda$ is longer than \codaiii{}'s box length. In order to estimate the mean free path along these lines of sight, we equate their high transmissions with the absence of large overdensities and assume that the density of neutral Hydrogen is constant and small along the LoS. Thus, we can obtain an estimate of the full mean free path by extrapolating the additional distance necessary for the integrated optical depth to reach one. Second, there are the cases of very opaque LoS (or at least starting in an opaque cell) where $\lambda$ is smaller than the cell size of \codaiii{}. Here, assuming that this cell does not contain an unresolved absorber, we use the cell's optical depth to deduce $\lambda$.

The left panel of Fig. \ref{fig:mfp_distrib_rahmati} shows the distributions for both the fitting method presented in \ref{sec:mthds}, and the method described above (that we call intercept method from hereon). Both methods are very consistent for short, long, and intermediate mean free paths, and throughout Reionization in \codaiii{}. However, there are some discrepancies, most notably for the shortest and longest mean free paths. This isn't too surprising as the exponential fit is harder in these cases, and at the same time we must make some strong assumptions for the intercept method. The right panel of Fig. \ref{fig:mfp_distrib_rahmati}, shows the average evolution of the ionised lines of sight as computed with all tested techniques, including the intercept method. As expected based on the distributions, the result is very similar to the evolution of the mean free path as derived using fitting. In fact, for $\rm z\leq 6$, the two methods are almost indistinguishable. Though at higher redshifts, the intercept method seems to predict a slightly higher mean free path. That these two methods independently yield such similar mean free paths, shows the robustness of our technique and results.

}


\bsp	
\label{lastpage}
\end{document}